\documentclass[conference]{IEEEtran}
\IEEEoverridecommandlockouts

\usepackage{cite}
\usepackage{amsmath,amssymb,amsfonts}

\usepackage{graphicx}
\usepackage{textcomp}
\usepackage[table]{xcolor}
\usepackage{xcolor}
\usepackage{tabulary}
\usepackage{array}
\usepackage{float}
\usepackage{makecell}
\usepackage{subfigure}
\usepackage[utf8]{inputenc}
\usepackage{ragged2e}
\usepackage[hidelinks]{hyperref}
\usepackage{authblk}

\usepackage{soul}
\usepackage{pifont}
\usepackage[table]{xcolor}
\usepackage{array}
\usepackage{ragged2e} 
\usepackage{placeins}
\usepackage{authblk}
\usepackage[utf8]{inputenc}
\usepackage[table]{xcolor}

\usepackage{algorithm}
\usepackage{algpseudocode}
\usepackage{amsmath}
\usepackage{amssymb}
\usepackage{geometry}
\DeclareUnicodeCharacter{202F}{\,}
\newcolumntype{P}[1]{>{\centering\arraybackslash}p{#1}}
\def\BibTeX{{\rm B\kern-.05em{\sc i\kern-.025em b}\kern-.08em
    T\kern-.1667em\lower.7ex\hbox{E}\kern-.125emX}}
    
\begin{document}

\title{Learning in Focus: Detecting Behavioral and Collaborative Engagement Using Vision Transformers\\
}

\author[1]{Sindhuja Penchala}
\author[1]{Saketh Reddy Kontham}
\author[1]{Prachi Bhattacharjee}
\author[1]{Nima Mahmoodi}
\author[1]{Daniel Fonseca}
\author[2]{\\Sareh Karami}
\author[2]{Mehdi Ghahremani} 
\author[2]{Andy D. Perkins}
\author[1]{Shahram Rahimi}
\author[1]{Noorbakhsh Amiri Golilarz}
\affil[1]{The University of Alabama, Tuscaloosa, AL, USA}
\affil[2]{Mississippi State University, Mississippi State, MS, USA}

\maketitle

\begin{abstract}
In early childhood education, accurately detecting collaborative and behavioral engagement is essential to foster meaningful learning experiences. This paper presents an AI-driven approach that leverages Vision Transformers (ViTs) to automatically classify children's engagement using visual cues such as gaze direction, interaction, and peer collaboration. Utilizing the Child-Play gaze dataset, our method is trained on annotated video segments to classify behavioral and collaborative engagement states (e.g., engaged, not engaged, collaborative, not collaborative). We evaluated six state-of-the-art transformer models: Vision Transformer (ViT), Data-efficient Image Transformer (DeiT), Swin Transformer, VitGaze, APVit and GazeTR. Among these, the Swin Transformer achieved the highest classification performance with an accuracy of 97.58\%, demonstrating its effectiveness in modeling local and global attention. Our results highlight the potential of transformer-based architectures for scalable, automated engagement analysis in real world educational settings.

\end{abstract}


\section{Introduction}

Today’s growing interest in AI-driven analytics has increased the importance of understanding behavioral and collaborative engagement, particularly in the context of early childhood \cite{Lu}. Understanding human behavior, which includes behaviors and interactions motivated by psychological, social, and environmental factors, is critical for encouraging collaborative engagement, in which individuals work together to achieve common goals. Such behaviors are critical for cognitive and social development: cooperatively engaged children demonstrate empathy, active communication, and cooperative problem solving, while gaze patterns such as eye contact and shared attention indicate key social and attentional states \cite{Chen}. In particular, gaze behavior serves as an important biomarker in developmental screenings, with aberrant gaze patterns typically associated with autism spectrum disorder(ASD), where difficulties with joint attention are among the most reliable early signs.

Student participation is widely considered an important indicator in education that influences the quality of knowledge construction and learning outcomes. High involvement indicates deep cognitive processing and sustained task time during learning activities, which is associated with improved understanding and academic performance \cite{Rogat}. Although early research focused on individual involvement, real-world classroom settings are essentially social: collaborative learning involves groups of students working together, and peer interactions can advance comprehension beyond isolated studies. In these environments, we can differentiate between behavioral participation (on-task activities and attentional focus) and collaborative participation (interactive participation and knowledge building among peers) \cite{Chen}. Understanding both forms of engagement is important because engaged students not only pay attention to assignments but also actively contribute to group learning, resulting in richer educational outcomes. 

Traditional engagement measurements, such as surveys or human observation, are labor intensive and coarse-grained \cite{Chen}. By assessing student's nonverbal signs in real-time, automated computer vision approaches provide a promising alternative. Vision-based systems use indicators such as facial expression, eye gaze, head posture, and body posture to determine attention and involvement \cite{Ai}. Prior works are divided into two categories: (1) feature-based models, which calculate hand-crafted signals (e.g., eye-gaze direction, facial action units, skeletal motions) and feed them into a classifier, and (2) end-to-end deep models, which learn engagement directly from raw video frames \cite{Ai}. Chen et al., combined gaze direction and facial expressions in a multi-modal network (MDNN) to predict collaborative learning engagement \cite{Chen}, while Abdelkawy et al., used a 3D CNN on upper body poses to recognize student actions and built a histogram of actions to classify on-task vs. off-task engagement \cite{Abdelkawy}.  These investigations demonstrate that both individual gaze cues and group interaction signals can be used automatically. However, manually combining several modalities can be complex and data-hungry, and many existing solutions still rely on significant annotation or simplified cases \cite{Lu}\cite{Ai}.

The latest advances in deep learning point to new prospects for engagement analysis. Vision Transformers (ViTs) have proven cutting-edge performance on image identification challenges by applying self-attention to picture patches. When pretrained on largescale datasets, ViTs can match or outperform CNNs while being extremely efficient \cite{Vit}. More broadly, the introduction of Large Vision Models (LVMs), which are similar to large language models, has introduced powerful foundational models for visual understanding \cite{Zhang24Yuchong}. These LVMs are trained on massive image collections to capture rich scene-level semantics, making them capable of handling difficult vision tasks and even multimodal reasoning \cite{Zhang24Yuchong}. Despite this promise, the application of ViT-based LVMs for classroom engagement has not been investigated. Most existing engagement detectors still rely on traditional CNNs or hybrid video-RNN. Thus, there is an opportunity to apply these modern vision architectures to better capture the spatiotemporal and attention dynamics of children’s interactions.

\begin{figure}
\centering
\includegraphics[width=0.45\textwidth]{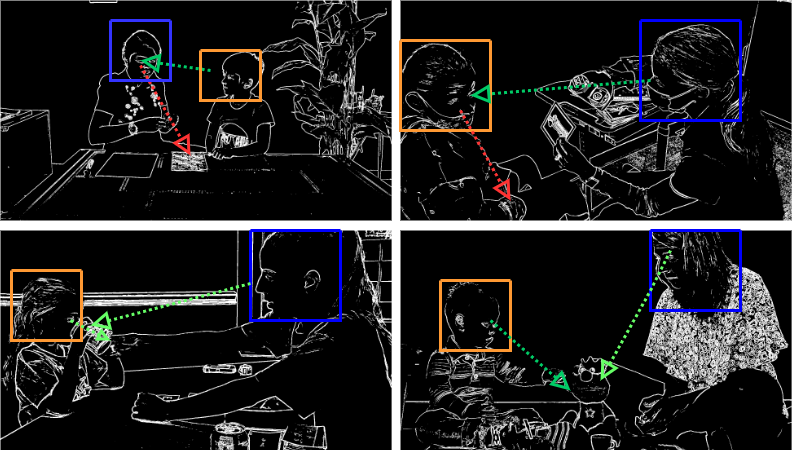}
\vspace{-0.3cm}
\caption{Edge-detected interaction samples of collaborative and behavioral engagement, highlighting gaze direction and participant bounding boxes. Blue and orange boxes represent adult and child participants, respectively. Green arrows indicate mutual or shared gaze inside the frame (engaged), while red arrows represent individual gaze directed outside the frame (not engaged), reflecting distinct engagement patterns.}\
\label{fig:Intro}
\end{figure}

To overcome these challenges, we introduced ``Learning in Focus'', a framework for identifying behavioral and collaborative involvement in children using video. We define behavioral engagement as a child's focused attention and on-task conduct during learning, and collaborative engagement as the level of interactive participation and joint problem solving within a group (see Fig. \ref{fig:Intro}. To illustrate model outputs while maintaining privacy and avoiding the exposure of identifiable features, we displayed only Canny edge-detected versions of the gaze frames in our figures. The original annotated images themselves were never shown or directly revealed). Using the ChildPlay gaze dataset \cite{dataset}, we process our model using a Vision Transformer that has been trained on large visual datasets. The primary aim of the project is to 
\begin{itemize}
 \item  Curate the ChildPlay gaze dataset by adding new
labels, and refining the labeling criteria.
    \item Automatically detect and classify children's behavioral and collaborative engagement (e.g., engaged or not engaged, collaborative or not collaborative) using the ChildPlay gaze video data.
    \item Evaluate and compare the effectiveness of several transformer-based architectures, such as ViT, Swin, and DeiT, VitGaze, APVit and GazeTR in classifying engagement and collaboration.
    \item Implement a Swin Transformer-based model that captures local cues and global context using multiscale patches and shifted window attention.
\end{itemize}

The rest of the paper is organized as follows. Section II presents related work on vision-based engagement detection and transformer models used in classroom settings. Section III describes the methodology, including the Swin Transformer architecture and its hierarchical components. Section IV discusses experimental setup, dataset preparation, evaluation metrics and compares the performance of Swin, ViT, DeiT, VitGaze, APVit and GazeTR models. Finally, Section V concludes the paper and outlines future directions for research in automated engagement recognition using visual data.

\section{Related works}

Vision-based engagement systems detect nonverbal cues and are used to infer attention, engagement, and collaboration. Recent research focuses on signs such as facial expressions, gaze, head and body posture, gestures, and spatial arrangement. For example, facial expression models (typically CNNs) are used to estimate emotional state (boredom, confusion, etc.), head-position or gaze trackers determine where students look, and full-body pose estimators identify behaviors. 


One hybrid technique named ``EngageSense" \cite{engageSense} trains a CNN on eye region pictures for gaze direction (99.5\% gaze accuracy) and uses OpenPose \cite{Tsai} for body points. Combining gaze and pose produces ~90\% accuracy in identifying students as fully or partially or not engaged \cite{engageSense}. CNNs (ResNet, MobileNet) extract facial/body features, which are then processed by classifiers or LSTMs to provide temporal context. Meta-learning has also been used in some studies. For example, Alarefah et al., \cite{alarefah2025} pre-trained a ViT on faces and added an LSTM (prototypical network) to handle few-shot student engagement classification, attaining state-of-the-art (SOTA)  on the EngageNet dataset \cite{singh2023}.

Recent vision based studies have investigated the automatic identification of student engagement via nonverbal clues. Wu \textit{et al.,}\cite{wu2024cmose} developed the CMOSE dataset, annotated by psychologists, to predict engagement using multimodal data. The study found that combining audio, video, and speech signals outperformed unimodal techniques. Fahid et al., \cite{fahid2023detecting} used facial video and chat logs to detect disengagement during middle school collaborative learning. They found that multimodal fusion enhances prediction accuracy.  Chen et al., \cite{chen2023mdnn} used a deep neural network to predict group participation by analyzing gaze and expression data. These works, together with datasets like DAiSEE \cite{gupta2016daisee} and RoomReader \cite{suzuki2023roomreader}, laid the framework for automated behavioral engagement tracking.

Transformer-based models have emerged as powerful tools for studying human behavior. Agrawal et al., \cite{agrawal2023forcedvit} developed a segmentation-guided Vision Transformer (ViT) for social behavior analysis that outperformed state-of-the-art benchmarks. Song \textit{etal.,}\cite{song2024vitgaze} used DINOv2 based ViTs for gaze prediction, which dramatically reduced model parameters while keeping good performance. Moreover, Suzuki \textit{et al.,}\cite{suzuki2025multistream} developed a global token attention model that incorporates multi-person audiovisual streams, resulting in accurate engagement estimates in group settings. 


\begin{figure*}
\centering
\includegraphics[width=1.0\textwidth]{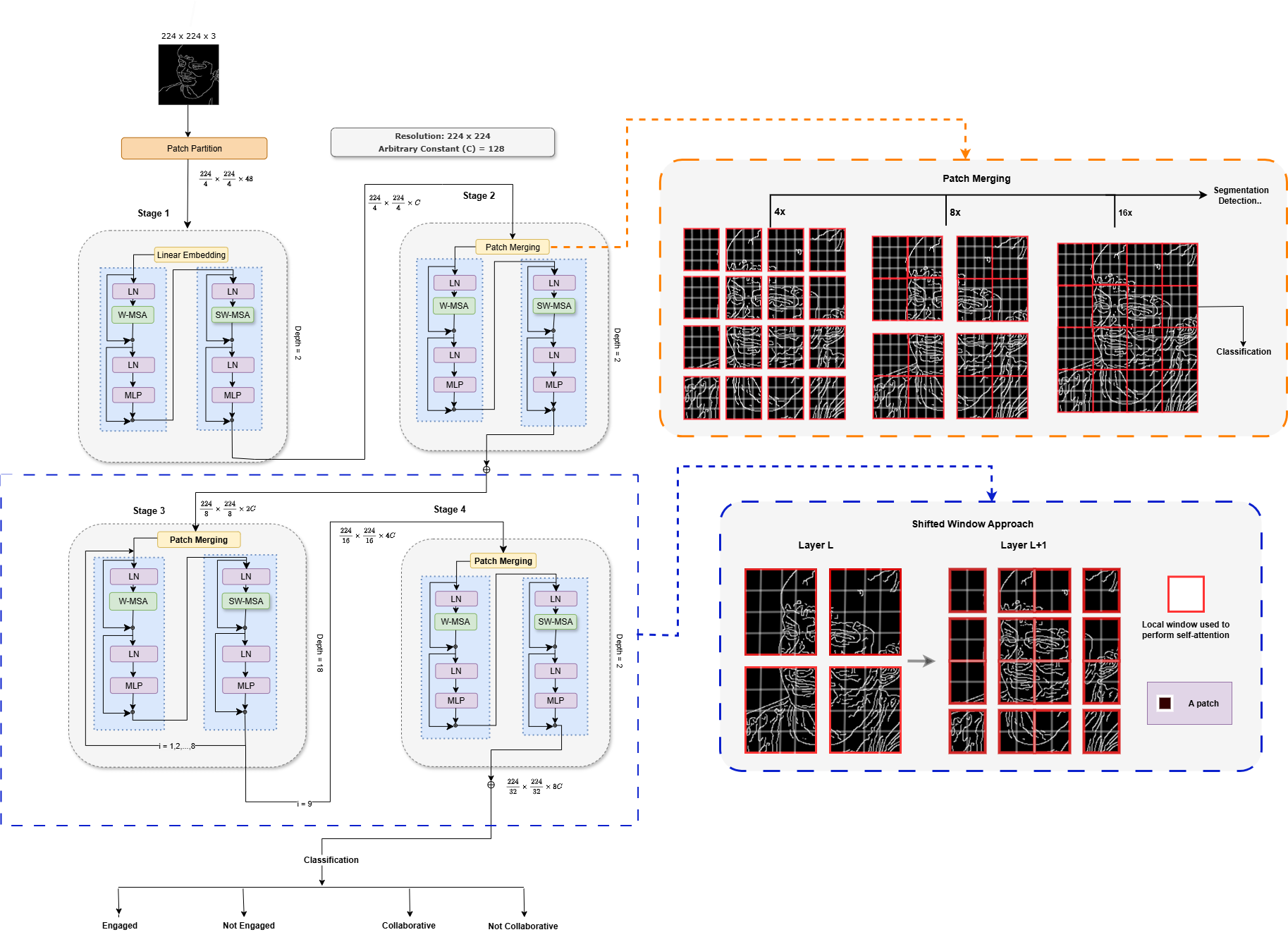}
\caption{Swin Transformer Base (Swin-B) architecture with a 2-2-18-2 block configuration across four stages. The input image is partitioned into 4×4 patches and linearly embedded to dimension C. Each stage consists of paired Swin Transformer blocks (i), alternating W-MSA and SW-MSA with LayerNorm, MLP, and residual connections. Patch Merging reduces spatial resolution and increases channel dimensions (C → 8C). The Shifted Window mechanism enhances cross-window attention \cite{Swin}. Final features are used for classification tasks such as engagement and collaboration detection.}\
\label{fig:SwinTrans}
\end{figure*}

Additionally, educational psychology defines engagement as a multidimensional construct\cite{fredricks2004engagement}, with behavioral engagement being a strong predictor of academic success\cite{reschly2008student}. Collaborative engagement emphasizes shared focus, coordination, and responsiveness among peers\cite{chen2023mdnn}. Recent research aligns these psychological definitions with computer vision features such as gaze, posture, and interaction sequences\cite{suzuki2025multistream}, shifting the measurement paradigm from self-reports to behaviorally grounded annotation. Despite advancement, obstacles persist.  Many data sets lack fine-grained frame-level labels \cite{suzuki2023roomreader}, which limits model generalization. Existing research often focuses on small groups, but real classrooms confront occlusion and scalability challenges\cite{beyan2023survey}. Deep models struggle with delicate behaviors and class imbalance, and few work addresses interpretability \cite{beyan2023survey}. Multimodal fusion is promising, but it is computationally demanding, which limits its practical application. Although vision and deep learning have improved engagement detection, important gaps remain in dataset quality, scalability, and model interpretability.  Our approach solves these issues using ViTs to provide a scalable and explainable framework for detecting collaborative and behavioral engagement in children via gaze data.

\section{Methodology}



\subsection{Background}
In 2020, Dosovitskiy et al., \cite{Vit} introduced a novel approach that applies a pure Transformer architecture, traditionally used in natural language processing, to visual data for image recognition tasks. The Vision Transformer (ViT), instead of using convolutional layers, divides each image into fixed-size patches, treats these patches as token sequences, and processes them using a standard Transformer encoder. This model surpasses traditional state-of-the-art convolutional neural networks (CNNs) when it is pre-trained on large-scale datasets such as ImageNet-21k or JFT-300M \cite{Russakovsky} \cite {JFT}. This demonstrates that the inductive biases that are inherent to CNNs are not strictly necessary if sufficient training data is available, thereby opening new directions for transformer-based models in computer vision tasks.

Building on this foundation, Liu et al., \cite{Swin} introduced the Swin Transformer, a hierarchical vision transformer architecture that includes two major innovations, which includes non-overlapping local windows and shifted window focus. Swin reduces complexity by limiting attention computation to small, localized windows, as compared to ViT, which applies global self-attention across all tokens. To ensure information flow across areas, it uses a shifted-window technique in alternating layers, which allows for interaction between nearby windows while maintaining efficiency. It also uses patch merging to create a hierarchical framework, allowing it to generate multi-scale feature representations similar to those found in CNNs.

This design makes it ideal for dense prediction tasks and scene understanding.  When tested on our engagement categorization challenge, the Swin Transformer outperformed all other models, with an accuracy of 97.58\%, indicating its better ability to predict both local and global relationships in visual input.

\begin{table*}[ht]
\centering
\caption{Stage-wise Configuration and Description of Swin-B (Base) Transformer \cite{Swin}.}
\vspace{-0.2cm}
\label{tab: stage}
\renewcommand{\arraystretch}{1.3}
\begin{tabular}{|p{1cm}|p{1.3cm}|p{1.8cm}|p{1cm}|p{1cm}|p{0.8cm}|p{0.8cm}|p{2.3cm}|}
\hline
\rowcolor{blue!20}
\textbf{Stage} & \textbf{Output Size} & \textbf{Patch Merging} & \textbf{Window Size} & \textbf{Embed Dim} & \textbf{Heads} & \textbf{Layers} & \textbf{Stage Description} \\
\hline
1 & $56 \times 56$ & Concat $4 \times 4$, 128-d, LN  & $7 \times 7$ & 128 & 4 & 2 & Splits image into patches and applies local attention. \\
\hline
2 & $28 \times 28$ & Concat $2 \times 2$, 256-d, LN & $7 \times 7$ & 256 & 8 & 2 & Reduces resolution and learns broader regional features. \\
\hline
3 & $14 \times 14$ & Concat $2 \times 2$, 512-d, LN & $7 \times 7$ & 512 & 16 & 18 & Deep blocks capture high-level global interactions. \\
\hline
4 & $7 \times 7$ & Concat $2 \times 2$, 1024-d, LN & $7 \times 7$ & 1024 & 32 & 2 & Forms final feature map for global representation. \\
\hline
\end{tabular}
\end{table*}


\begin{algorithm}
\caption{Swin transformer based behavioral and collaborative engagement detection.}
\label{alg: Swinalgo}
\begin{algorithmic}[1]
\Require RGB image of size 224×224
\State divide the image into non-overlapping $4 \times 4$ patches
\State convert each patch into a feature vector using a linear projection
\State form a 2D grid of patch embeddings
 
\For{each block in Stage 1}
    \State apply window-based self-attention within $7 \times 7$ regions
    \State alternate attention with shifted windows for cross-region interaction
    \State apply MLP with skip connection
\EndFor
 
\State merge every $2 \times 2$ patch group to reduce spatial size and increase depth
 
\For{each block in Stage 2}
    \State repeat attention + shifted windows and MLP with skip connections
\EndFor
 
\State merge patches again for Stage 3
 
\For{each block in Stage 3}
    \State repeat the attention and MLP process
\EndFor
 
\State merge patches again for Stage 4
 
\For{each block in Stage 4}
    \State apply attention globally (entire $7 \times 7$ grid fits in one window)
    \State apply final MLP block
\EndFor
 
\State apply global average pooling to reduce $7 \times 7$ grid to a single vector
\State pass the vector through the final classification layer
\State apply \textbf{argmax} to select the highest scoring class
\State map index to class label from \{``engaged'', ``not\_engaged'', ``collaborative'', ``not\_collaborative''\}
\State \Return predicted class label
\end{algorithmic}
\end{algorithm}

\subsection{Method}

Our proposed model is built upon a four-stage Swin Transformer architecture \cite{Swin} designed for vision-based behavioral classification using only static RGB image input. It operates on 224×224 images and performs multi-class classification across four classes: Engaged, Not Engaged, Collaborative, and Not Collaborative. The network follows the hierarchical structure of the Swin Transformer, consisting of four sequential stages that gradually reduce the spatial resolution of feature maps while increasing their channel dimensions to construct rich, multi-scale representations. The final output is passed through a single classification head based on a Multi-Layer Perceptron (MLP) that predicts one of the four target categories.

Fig \ref{fig:SwinTrans} illustrates the architectural workflow of the Swin Transformer used for engagement and collaboration classification. The model begins by partitioning an input image of size $224 \times 224 \times 3$ into non-overlapping $4 \times 4$ patches, each of which is linearly embedded into a lower-dimensional representation. This is followed by four hierarchical stages, each consisting of multiple Transformer blocks and a patch merging layer. At each stage, the spatial resolution is progressively reduced (e.g., $56 \times 56$ to $7 \times 7$), while the embedding dimension is increased (e.g., from 128 to 1024), enabling the model to efficiently aggregate local features into more abstract and semantically rich representations (see Table \ref{tab: stage}. It describes the stage-wise architecture of the Swin-B (Base) Transformer, highlighting how image features are progressively refined).

To maintain both local context and global connectivity, the architecture alternates between standard window-based multi-head self-attention (W-MSA) and shifted window-based self-attention (SW-MSA) mechanisms. These are shown in Layer $l$ and Layer $l+1$ of the diagram (on the bottom right), where attention is initially restricted to fixed-size local windows, and then the windows are shifted in the following layer to allow cross-window information flow. This shift ensures that previously unconnected tokens can now interact, leading to enhanced global feature learning.

Additionally, the figure on the right visually demonstrates the multiscale patch merging strategy. The image is shown at different levels of downsampling (4×, 8×, 16×), where early layers focus on fine-grained local details, and deeper layers capture larger contextual regions. The resulting features from the final stage are passed to a shared MLP head for multi-label classification across different categories of Engaged, Not Engaged, Collaborative, and Not Collaborative. The inference procedure applied during the classification of collaborative and behavioral engagement is detailed in Algorithm \ref{alg: Swinalgo}.

\section{Experimental Results and Discussion}

This section presents and analyzes the experimental results obtained from training and evaluating the ViT \cite{Vit}, DeiT \cite{Deit}, VitGaze \cite{vitgaze}, APVit \cite{apvit} and GazeTR \cite{gazetr} and Swin \cite{Swin} Transformer models, focusing on their performance across various metrics, learning behaviors, and classification tasks.


\subsection{Experimental Setup}

\subsubsection{Data Preparation}

We used the ChildPlay Gaze dataset \cite{dataset} from the Zenodo repository as the data source. It comprises of 401 short video clips, extracted from 95 videos showing children engaged in free play and interaction with adults in uncontrolled environments (like kindergartens, preschools, therapy centers). The videos are primarily shot indoors and feature at least one adult with 1 to 2 children playing with toys or participating in exercises.


Using this dataset, we labeled the images based on the criteria in Table \ref{tab:engagement_criteria}, which defines labeling rules for four engagement categories and includes brief explanations to ensure consistency during model training. The dataset curated in this study consists a total of 4,538 samples distributed across four categories: engaged (2,793), not engaged (545), collaborative (826), and not collaborative (374). As observed, the class distribution is highly imbalanced, with a significant majority of samples belonging to the engaged category.

 
To address this imbalance and improve model generalization,  stratified 5-fold cross-validation and data augmentation technique was applied to the training set. Stratified 5-fold cross-validation was applied to ensure balanced class representation across folds. However, this approach did not yield better results and showed a noticeable drop in training accuracy compared to the standard train-test split. This suggests that the model struggled to maintain consistency across folds, possibly due to the class imbalance and limited sample size in minority classes.

Additionally, we employed data augmentation through oversampling to mitigate the effects of class imbalance. This technique involved duplicating samples from the minority classes (``Non Collaborative") to match the number of instances in the majority classes (``Engaged"). By increasing the representation of underrepresented classes, the model was better exposed to their patterns during training. This approach helped to reduce bias towards the majority classes and improved the model’s ability to generalize across all categories of behavior and collaboration.
 
 Specifically, the following transformations were used: 

\begin{itemize}
    \item \textbf{Random Resized Crop:} Applied with a scale range of (0.7, 1.0) to simulate varied object sizes and framing.
    \item \textbf{Random Horizontal Flip:} Used to introduce left-right orientation variability in the images.
    \item \textbf{Random Rotation:} Images were randomly rotated up to 15 degrees to improve rotational invariance.
    \item \textbf{Color Jitter:} Brightness, contrast, saturation, and hue were randomly adjusted to mimic lighting and color changes.
    \item \textbf{Normalization:} Images were normalized using predefined mean and standard deviation values.
\end{itemize}

\subsubsection{Metrics}

To evaluate the performance of the proposed model, we used various sets of metrics, including accuracy, precision, recall, F1 score, ROC curve, and confusion matrix. We have also plotted the graphs of the learning curve and training logs to check how best the models are performing in detecting the engagement status of the child.

\begin{table*}[ht]
\centering
\small
\caption{Labeling Criteria for Behavioral and Collaborative Engagement in ChildPlay Gaze Dataset.}
\vspace{-0.2cm}
\begin{tabular}{|>{\centering\arraybackslash}m{3.5cm}|
                >{\justifying\arraybackslash}m{7.5cm}|
                >{\centering\arraybackslash}m{4.3cm}|}
\hline
\rowcolor{blue!20}
\textbf{Label} & \centering\textbf{Description} & \textbf{Sample Images} \\
\hline

\makecell{\textbf{Engaged} \\ \textbf{(attentive to the activity)}} &
\noindent The child is clearly interested in a specific on-screen target, interacting with the environment. This is an appropriate focus, hence engaged. 
& \includegraphics[scale=0.286]{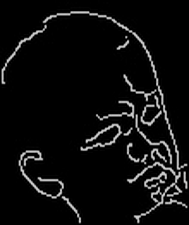} \includegraphics[width=1.9cm]{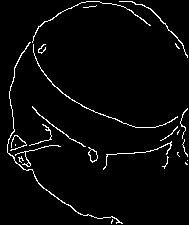}
\\
\hline
\makecell{\textbf{Not Engaged} \\ \textbf{(distracted)}} &
\noindent The child’s attention is away from the on-screen activity, indicating disengagement. They are not interacting with anything we care about in the scene, so they’re not engaged at this moment. & \includegraphics[scale=0.142]{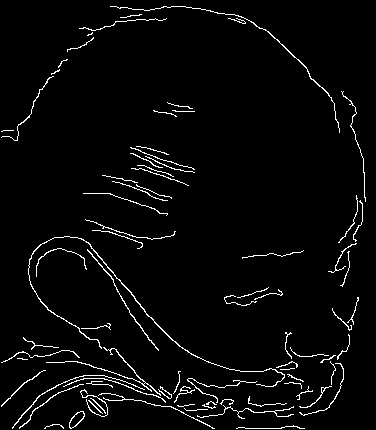} \includegraphics[width=2.002cm]{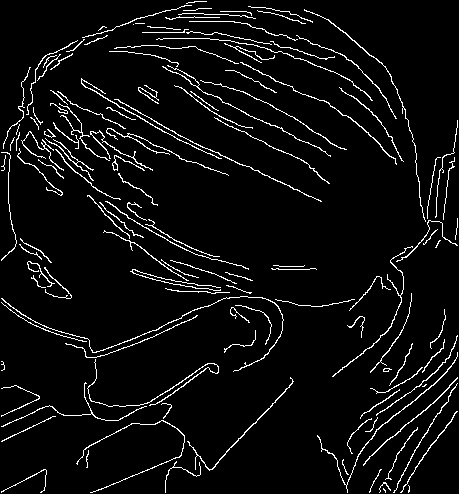}\\
\hline
\makecell{\textbf{Collaborative} \\ \textbf{(socially engaged)}} &
\noindent These behaviors show joint attention and interaction. The child is involved with both the task and the partner. Looking back-and-forth between a toy and adult indicates the child is including the adult in their play or seeking communication (a clear sign of collaboration).  & \includegraphics[scale=0.246]{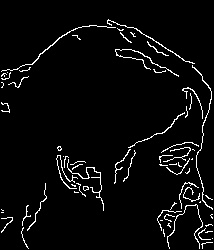} \includegraphics[width=2.0cm]{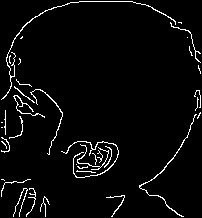}\\
\hline
\makecell{\textbf{Not Collaborative} \\ \textbf{(engaged but not socially)}} &
The child is engaged with the activity but not interacting with the partner. They show no shared attention. Even though they are focused (engaged), they are not collaborating. & \includegraphics[scale=0.171]{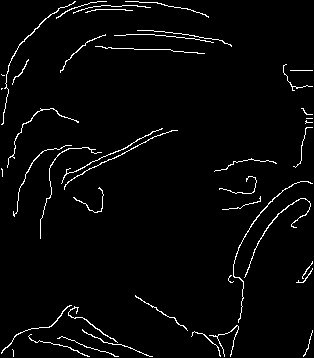} \includegraphics[width=2.0cm]{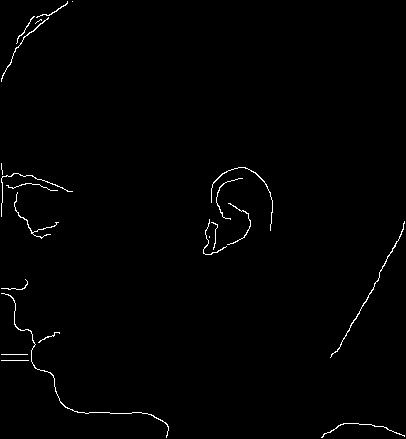}\\
\hline
\end{tabular}
\label{tab:engagement_criteria}
\end{table*}


\subsubsection{Hyperparameters}

To enable an accurate comparison, all six models (ViT, Swin, VitGaze, APVit and GazeTR and DeiT) were trained under the same parameters. As can be seen from Table I, all models employed the same learning rate of 2e-4, batch size of 32, and weight decay of 1e-2. Each model was trained with the AdamW optimizer, which is noted for its stability and effectiveness when training transformer architectures. To improve model generalization and address class imbalance, a shared set of data augmentation strategies was used. These included random resizing cropping, horizontal flipping, random rotation, color jittering, tensor conversion, and normalizing. By using similar hyperparameters and augmentation techniques, the evaluation focuses solely on architectural differences, guaranteeing that any performance variances detected are due to model capabilities rather than tuning inconsistencies.

\newcommand{\bestcell}[1]{\cellcolor{green!30}#1}
\begin{table}[ht]
\centering
\caption{Performance Comparison of Transformer Models.}
\vspace{-0.2cm}
\begin{tabular}{|p{0.9cm}|p{1.0cm}|p{0.7cm}|p{0.7cm}|p{1.0cm}|p{1.0cm}|}
\hline
\rowcolor{blue!20}
\textbf{Model} & \textbf{Precision} & \textbf{Recall} & \textbf{F1 Score} & \textbf{Accuracy} & \textbf{Params} \\
\hline
ViT & 0.9495 & 0.9675 & 0.9583  & 0.9725 & 86M\\
\hline
DeiT & \bestcell{\textbf{0.9623}} & 0.9710 & 0.9662 & 0.9747 & 86M\\
\hline
VitGaze & 0.7998 & 0.7984 & 0.7945 &  0.8502 & 22M\\
\hline
Apvit & 0.9323 & 0.9679 & 0.9482 &  0.9593 & - \\
\hline
GazeTR & 0.9244 & 0.9709 & 0.9460 &  0.9570 & - \\
\hline
\textbf{Swin} & 0.9611 & \bestcell{\textbf{0.9805}} & \bestcell{\textbf{0.9701}}& \bestcell{\textbf{0.9758}} & 88M\\
\hline
\end{tabular}
\label{tab:transformer_performance}
\end{table}


\subsection{Results}

We conducted experiments to compare the performance of six cutting-edge transformer-based models: ViT \cite{Vit}, Swin \cite{Swin}, DeiT \cite{Deit}, VitGaze \cite{vitgaze}, APVit \cite{apvit} and GazeTR \cite{gazetr} using a labeled dataset consisting of 4,537 tagged images. All the models were fine tuned with pretrained weights derived from large scale image datasets. It allows us to use rich feature representations while achieving robust performance on a relatively small dataset.


\begin{figure*}
  \centering
  \subfigure[ViT]{\includegraphics[width=0.3\textwidth]{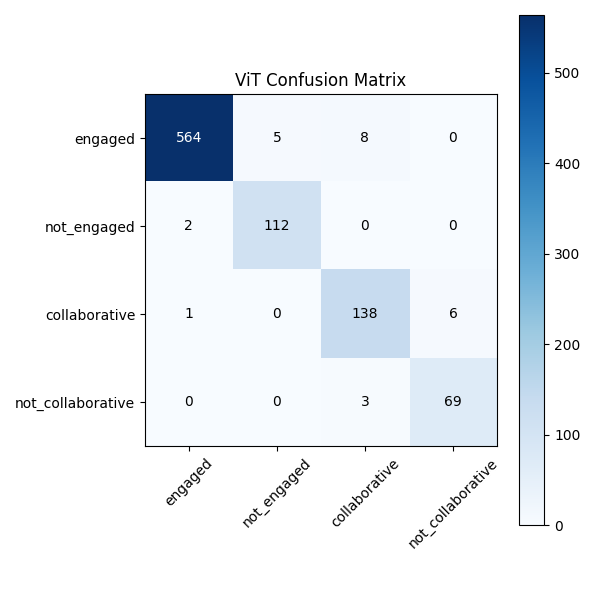}} 
  \subfigure[DeiT]{\includegraphics[width=0.3\textwidth]{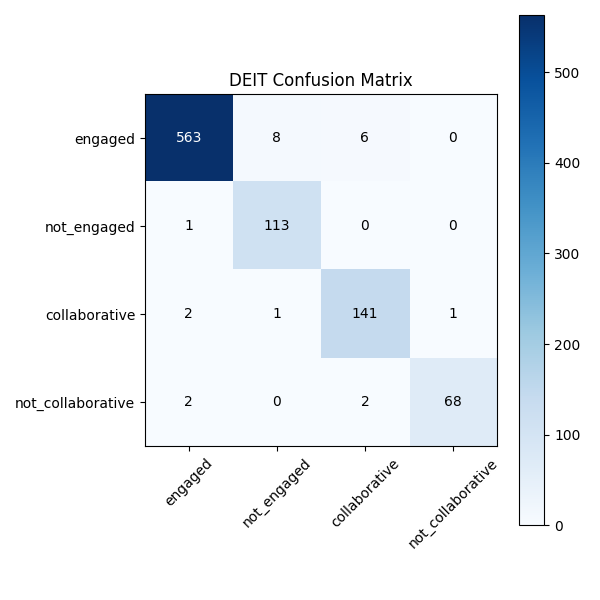}}  
    \subfigure[ViTGaze]{\includegraphics[width=0.3\textwidth]{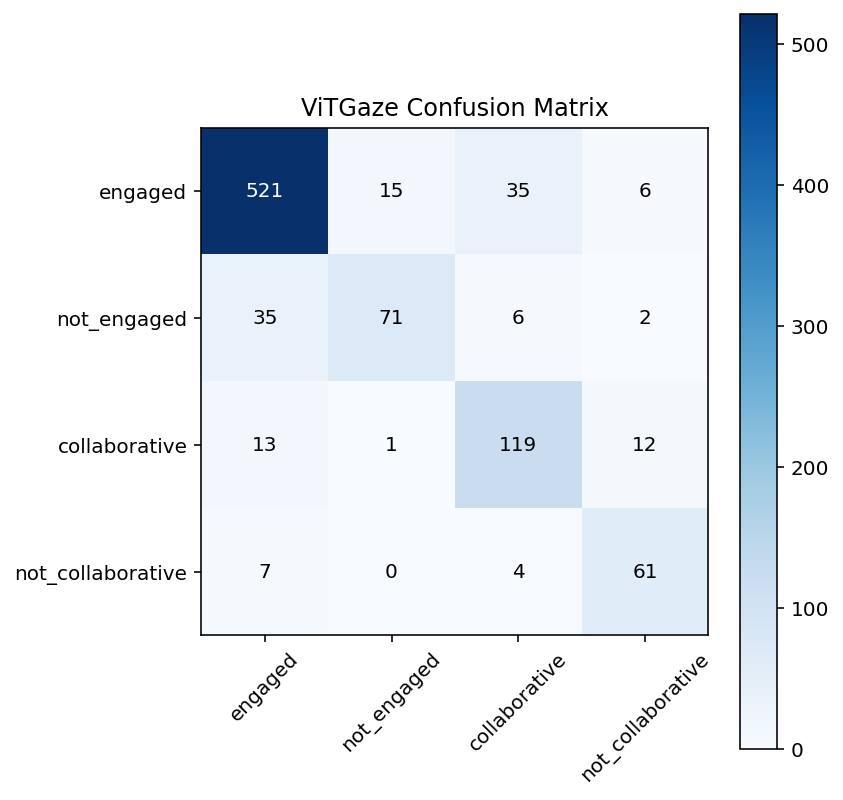}} \\
  \subfigure[APvit]{\includegraphics[width=0.3\textwidth]{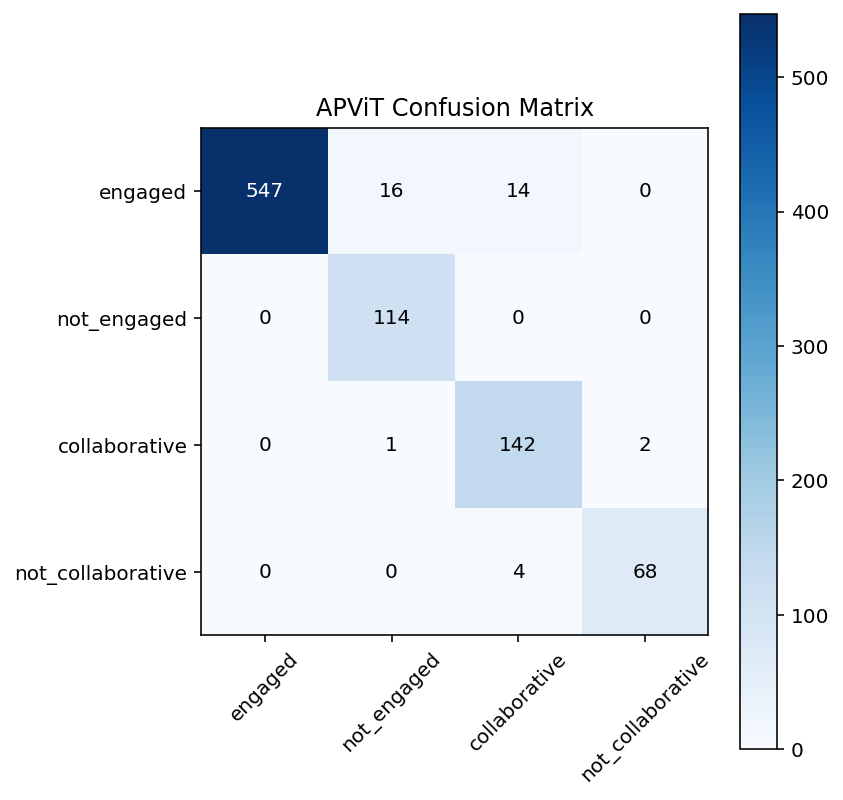}} 
  \subfigure[GazeTR]{\includegraphics[width=0.3\textwidth]{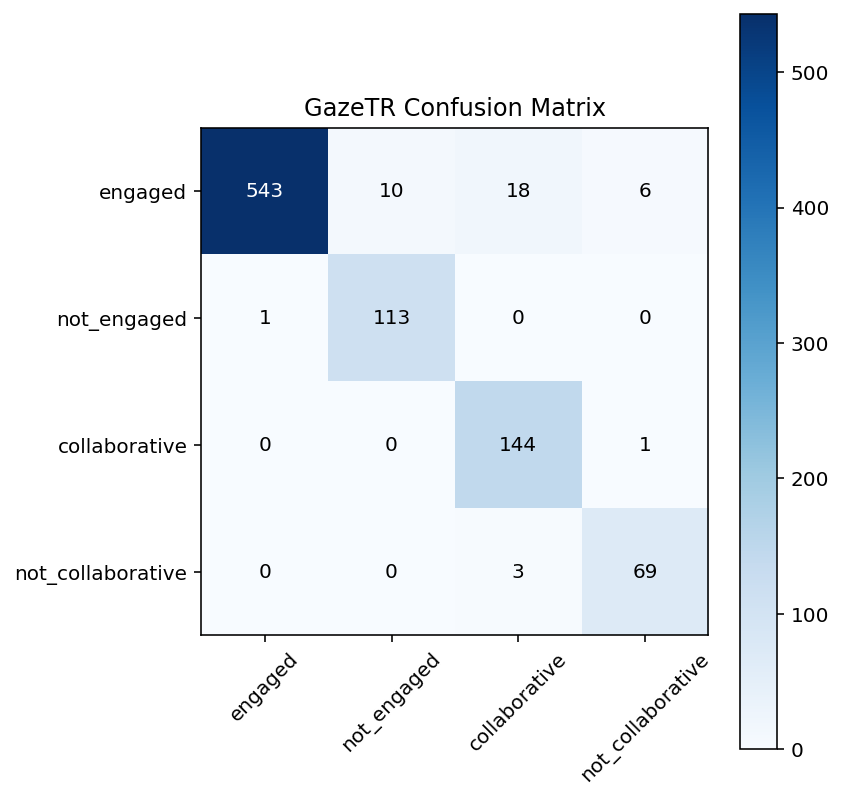}}
    \subfigure[Swin]{\includegraphics[width=0.3\textwidth]{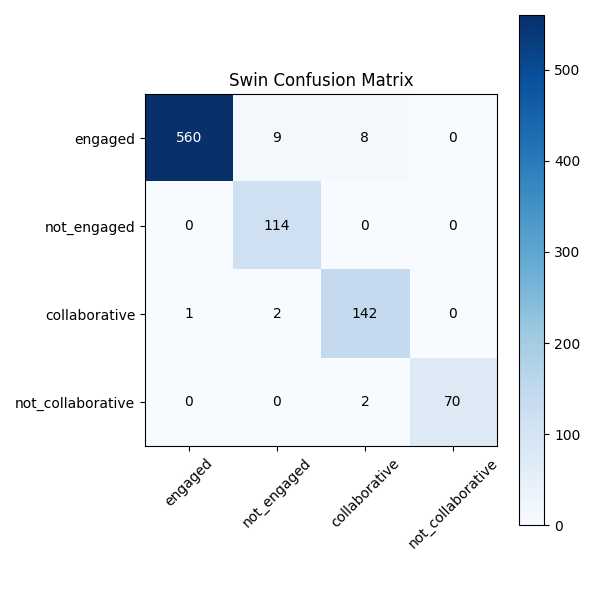}}\\
  \caption{Confusion matrices of different transformer models where rows correspond to actual labels and columns to predicted labels.}
  \label{fig: Confusion Matrices of Transformer Models}
\end{figure*}

\begin{figure*}
  \centering
  \subfigure[ViT]{\includegraphics[width=0.3\textwidth]{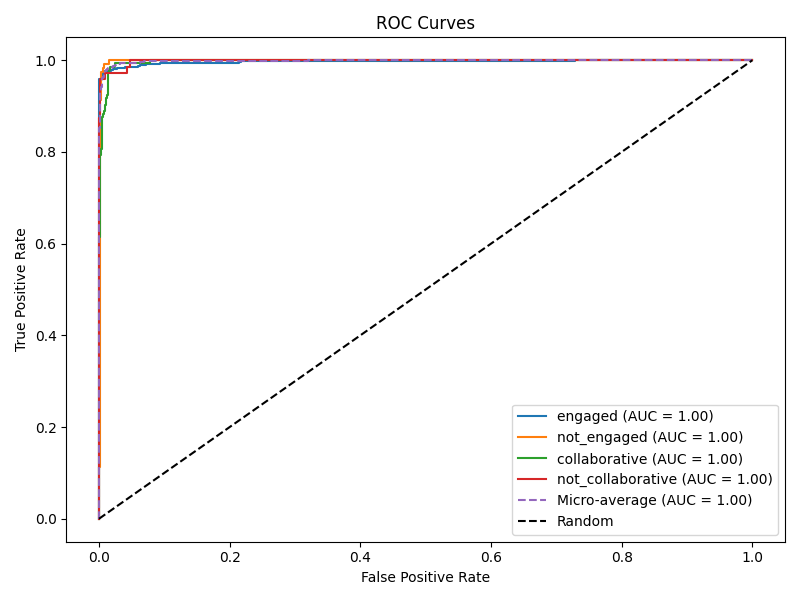}} 
  \subfigure[DeiT]{\includegraphics[width=0.3\textwidth]{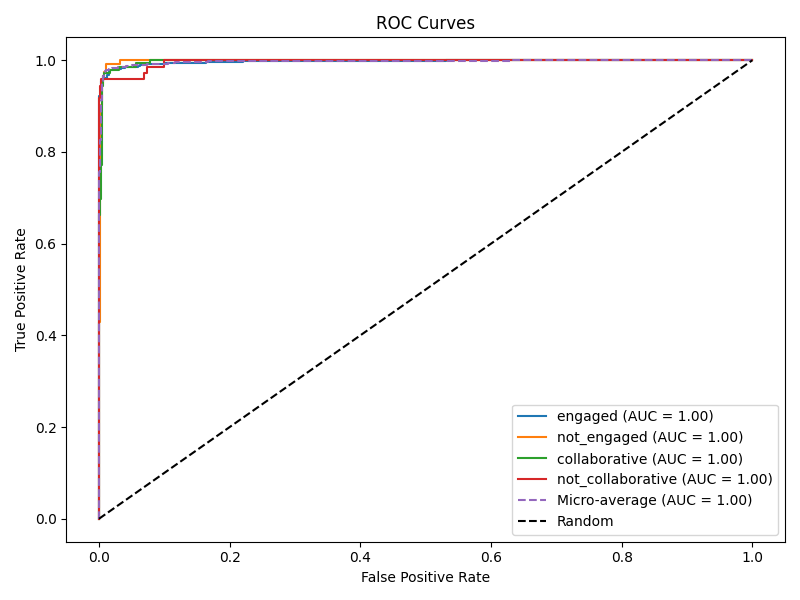}} 
  \subfigure[ViTGaze]{\includegraphics[width=0.3\textwidth]{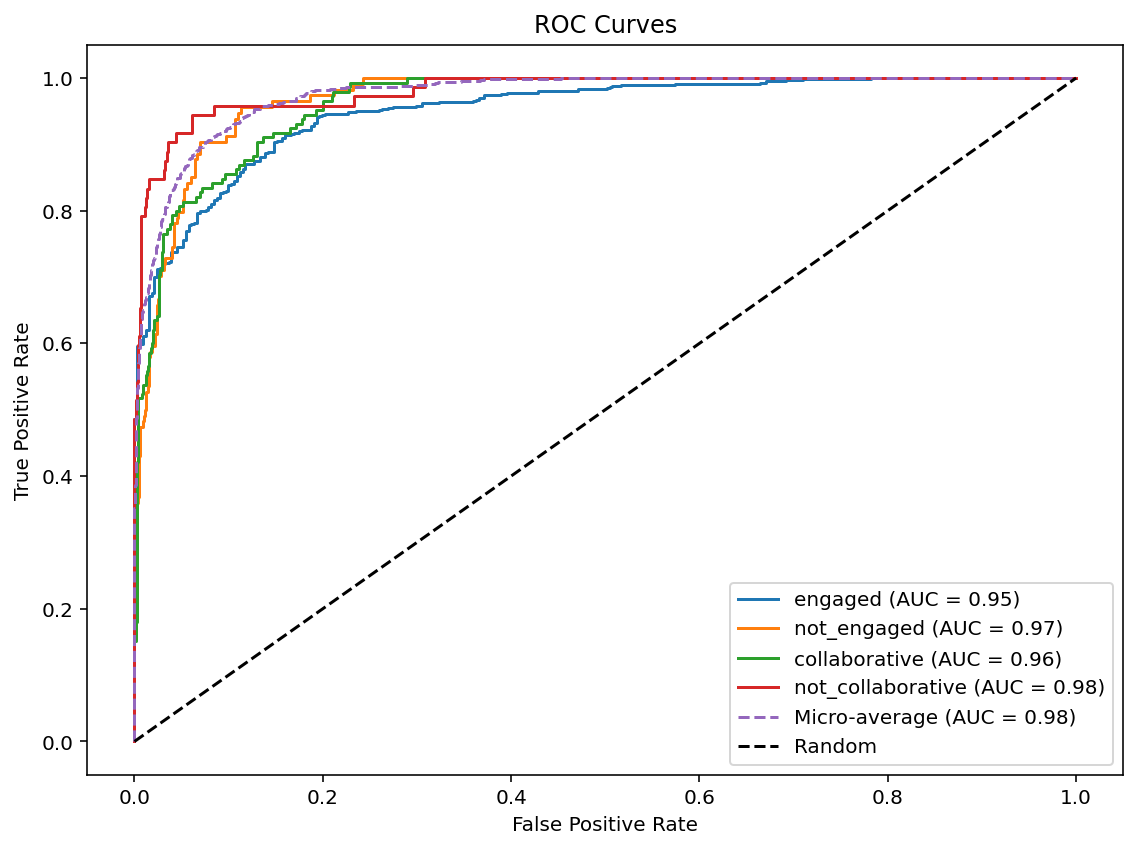}} 
  \subfigure[APVit]{\includegraphics[width=0.3\textwidth]{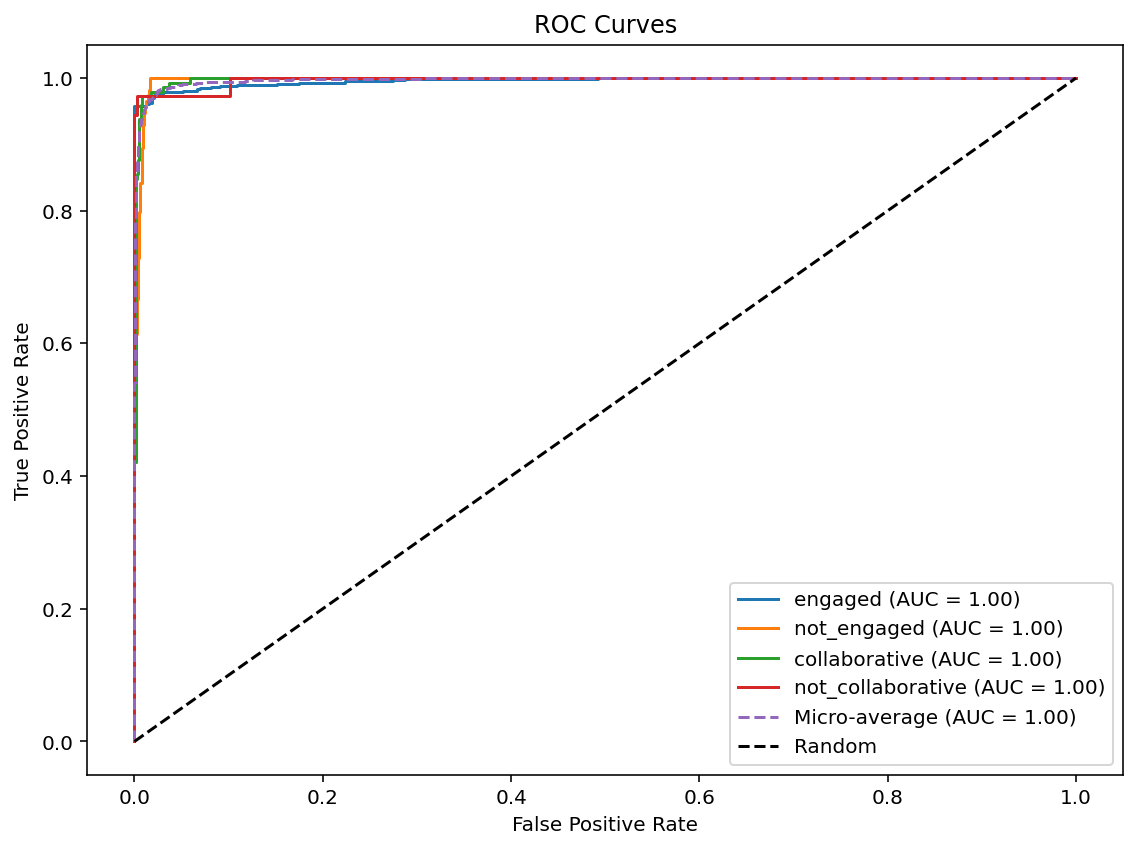}} 
  \subfigure[GazeTR]{\includegraphics[width=0.3\textwidth]{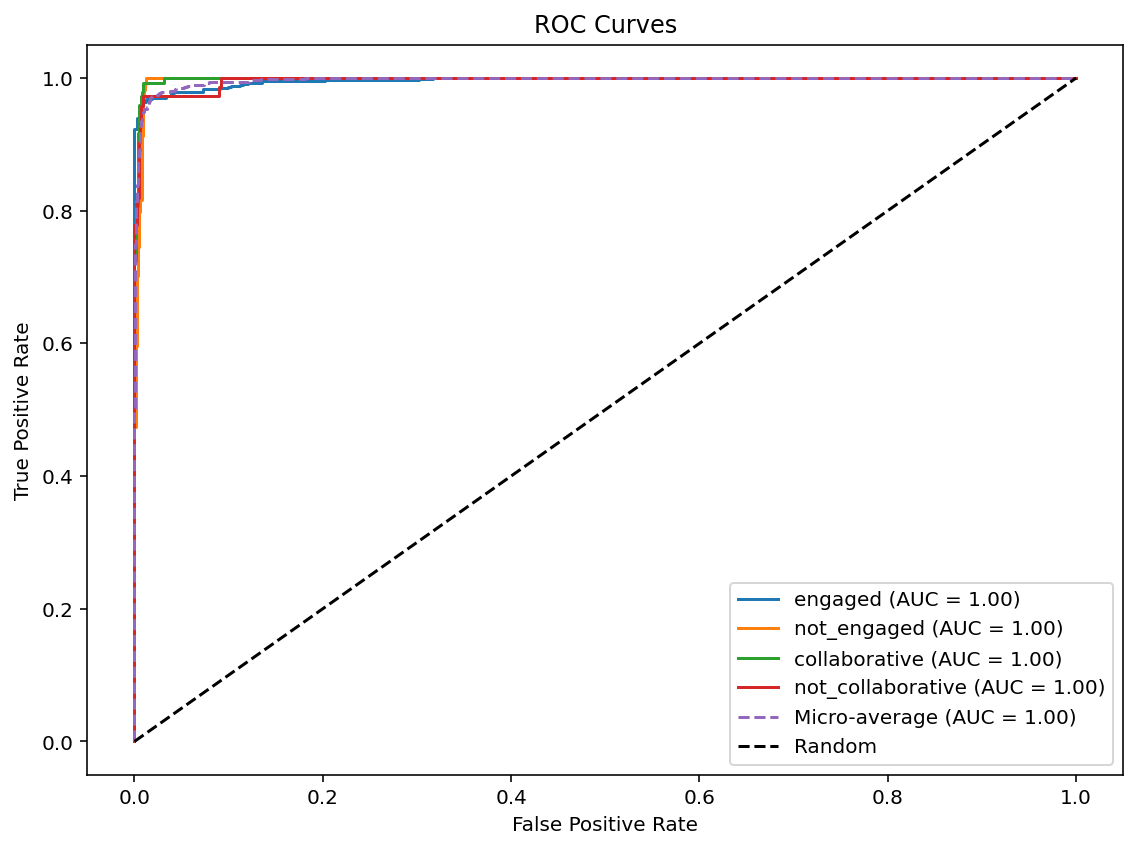}}
    \subfigure[Swin]{\includegraphics[width=0.3\textwidth]{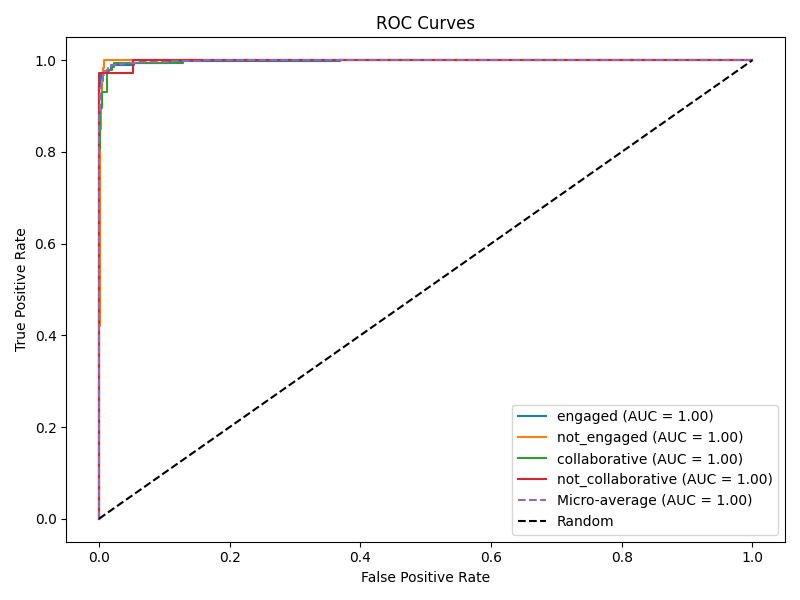}}
  \caption{ROC curves of different transformer models.}
  \label{fig: ROC Curves of Transformer Models}
\end{figure*}



\subsubsection{Statistical Comparison}

Table \ref{tab:transformer_performance} presents a comparative evaluation of six transformer-based architectures: ViT, DeiT, VitGaze, ApViT, GazeTR, and Swin across four key performance metrics. Among all models, the Swin Transformer achieves the highest overall performance, obtaining the top recall (0.9805), F1-score (0.9701), and accuracy (0.9758), demonstrating strong capability in capturing fine-grained engagement cues. DeiT records the highest precision (0.9623), highlighting its ability to minimize false positives while maintaining accuracy comparable to ViT. The standard ViT model also performs strongly, achieving an F1-score of 0.9583, indicating that baseline transformer encoders remain highly competitive.

ApViT shows balanced and reliable performance with an F1-score of 0.9482, positioning it between lightweight gaze-specific models and larger transformer variants. Gaze-focused models such as VitGaze and GazeTR also perform well. However, GazeTR demonstrates notably stronger performance, achieving a high recall of 0.9709 and an accuracy of 0.9570, which is substantially higher than the accuracy of VitGaze (0.8502). Overall, the results indicates that while several models perform competitively, hierarchical designs such as Swin offer the best overall balance of accuracy and robustness for holistic engagement recognition.

\subsubsection{Confusion Matrix}

Fig \ref{fig: Confusion Matrices of Transformer Models} (a)–(f) show the confusion matrices for the ViT, DeiT, VitGaze, APVit, GazeTR and Swin Transformers highlighting each model's ability to classify engagement and collaboration levels. All models successfully identified the majority of samples across classes, with the Swin Transformer producing the most balanced and accurate results. ViT performed well overall, but there was some confusion between the ``engaged" and ``collaborative" classes. DeiT showed better precision, particularly in the ``collaborative" category, despite minor misclassifications across all classes. 

Although ViTGaze benefits from a lightweight design, its confusion matrix shows comparatively higher misclassification, particularly between engaged, not engaged, and collaborative states. In contrast, APViT and GazeTR exhibit much cleaner class separability, with strong diagonal dominance, while Swin delivers the most balanced and consistent predictions overall.

\subsubsection{ROC Curve}

Fig \ref{fig: ROC Curves of Transformer Models} (a)–(f) display the ROC curves for the six transformer-based models across the four behavioral categories. Five of the models (ViT, DeiT, APViT, GazeTR, and Swin) achieve near-perfect discriminative performance, with AUC values approaching 1.00 across all classes. ViTGaze shows slightly lower AUC values (0.95–0.98), indicating reduced sensitivity and specificity compared to the others. Although all models produce clustered curves near the top-left region (indicating low false-positive rates and high true-positive rates), Swin and APViT exhibit the most stable and consistently smooth ROC profiles, reflecting stronger confidence across decision thresholds.

\section{Conclusion}

In this paper, we explored how transformer based models can be used to detect behavioral and collaborative engagement in learning environments. Using cropped images extracted from the ChildPlay Gaze dataset, we trained and evaluated models like ViT, DeiT, VitGaze, APVit, GazeTR and Swin Transformer to recognize different types of engagement. Among them, the Swin Transformer stood out for its ability to capture both fine-grained and broad visual patterns, leading to the best overall performance.
In the next phase, we plan to advance from static image classification to a more dynamic video-based understanding. By incorporating transformer-based models like the Video Swin Transformer, we aim to effectively capture temporal patterns such as gaze shifts and interaction cues over time. This will enable behavior classification at the frame level and support real-time analysis as well. Overall, this pipeline will help us build a context-aware and scalable system that better aligns with real-world scenarios and improves the automated detection of engagement and collaboration.

\vspace{0.2cm}

\noindent\textbf{Acknowledgement:}
This work was supported by the BRAINS Lab (Bioinspired Robotics, AI, Imaging \& Neurocognitive Systems Laboratory) in the Department of Computer Science at The University of Alabama.


\end{document}